\documentclass[useAMS,usenatbib]{mn2e}
\usepackage{graphicx}
\title[Global $21$~cm: chromatic effects]
  {Chromatic effects in the $\mathbf{21}$~cm global signal from the cosmic~dawn}
\author[Vedantham et al.]
  {H.K.~Vedantham$^{a}$\thanks{E-mail: harish@astro.rug.nl}, L.V.E.~Koopmans$^{a}$, A.G.~de~Bruyn$^{a,b}$, S.J.~Wijnholds$^{b}$, \newauthor B.~Ciardi$^{c}$, and M.A.~Brentjens$^{b}$\\
$^{a}$ Kapteyn Astronomical Institute, University of Groningen, P.O. Box 801, 9700 AV Groningen, The Netherlands\\
$^{b}$Netherlands Institute for Radio Astronomy (ASTRON), Oude Hoogeveensedijk 4, 7991 PD Dwingeloo, The Netherlands\\
$^{c}$ Max-Planck Institute for Astrophysics, Karl-Schwarzschild-Strasse 1, 85748 Garching bei M\"{u}nchen, Germany}
\begin{document}
%
%\date{Accepted unknown. Received unknown; in original form unknown}
\date{\today}
\pagerange{\pageref{firstpage}--\pageref{lastpage}} \pubyear{2012}
\def\LaTeX{L\kern-.36em\raise.3ex\hbox{a}\kern-.15em
    T\kern-.1667em\lower.7ex\hbox{E}\kern-.125emX}
\newtheorem{theorem}{Theorem}[section]
\label{firstpage}
\maketitle
%
%                             ___________
%                            /           \
% ----------------------------- ABSTRACT --------------------------
%                            \___________/
%

\begin{abstract}
The redshifted $21$~cm brightness distribution from neutral hydrogen is a promising probe into the cosmic dark ages, cosmic dawn, and re-ionization. LOFAR's Low Band Antennas (LBA) may be used in the frequency range $45$~MHz to $85$~MHz ($30>z>16$) to measure the sky averaged redshifted $21$~cm brightness temperature as a function of frequency, or equivalently, cosmic redshift. These low frequencies are affected by strong Galactic foreground emission that is observed through frequency dependent ionospheric and antenna beam distortions which lead to chromatic mixing of spatial structure into spectral structure. Using simple models, we show that (i) the additional antenna temperature due to ionospheric refraction and absorption are at a $\sim 1$\% level--- $2$ to $3$ orders of magnitude higher than the expected $21$~cm signal, and have an approximate $\nu^{-2}$ dependence, (ii) ionospheric refraction leads to a knee-like modulation on the sky spectrum at $\nu\approx 4\times$plasma frequency. Using more realistic simulations, we show that in the measured sky spectrum, more than $50$\% of the $21$~cm signal variance can be lost to confusion from foregrounds and chromatic effects. We conclude that foregrounds and chromatic mixing may not be subtracted as generic functions of frequency as previously thought, but must rather be carefully modeled using additional priors and interferometric measurements.  
\end{abstract}
%
% ABSTRACT SCRATCH PAD
% Besides large synthesis arrays (like LOFAR, MWA, PAPER, and GMRT) that are attempting to measure the angular power spectrum of $21$~cm brightness fluctuations, 
%In this paper we not only quantify these chromatic effects in global $21$~cm experiments, but also quantify to what extent they confuse the $21$~cm signature. 
%
%                             ___________
%                            /           \
% --------------------------- INTRODUCTION --------------------------
%                            \___________/
%

\section{Introduction}
\label{sec:intro}
Neutral Hydrogen (H\textsc{i}) interacts with $21$~cm photons through a spin-flip transition \citep{hulst45}. Observing the redshifted $21$~cm brightness temperature\footnote{Throughout this paper, when we say $21$~cm signal or $21$~cm brightness temperature we, really mean \emph{redshifted} $21$~cm signal} against a background of the cosmic microwave background is a promising tracer of the cosmic dark ages, cosmic dawn, and the epoch of reionization \citep{field59,madau97,sun72,sun75}. Detecting the spatial fluctuations of $21$~cm brightness requires many hundreds of hours of integration with large radio synthesis telescopes, owing to its faintness as compared to Galactic and Extragalactic foregrounds \citep{lofarsens,mwasens,papersens}. On the other hand, the sky averaged $21$~cm brightness--- also called the global signal, is bright enough to be measured within a day's worth of integration based on a signal to noise ratio argument \citep{shaver99}. Since the received frequency of redshifted $21$~cm photons corresponds to cosmic redshift, accurately estimating the sky averaged brightness temperature as a function of frequency will provide insights into the evolution of H\textsc{i} during the dark ages, cosmic dawn, and the epoch of reionization \citep{sethi05,furlanetto06b,prit10}.\\

Thermal uncertainties are not the limiting factor in global $21$~cm experiments, and spectral contamination due to systematic artifacts have impeded a reliable detection thus far \citep{chip09,bowman08}. In particular, since the signal in such experiments is the variation of $21$~cm brightness temperature with frequency, any instrumental or observational systematic that affects the measured bandpass power poses a severe limitation. These systematics are especially limiting since the foregrounds are $\sim 5$ orders of magnitude larger than the expected $21$~cm signal. Consequently, measuring the $21$~cm signal spectrum requires precise understanding of frequency dependent effects of instrumental gain, instrumental noise contribution, antenna beam shape, and ionospheric effects, coupled with spatially and spectrally varying foregrounds (Galactic and Extragalactic). The effects of these parameters are not always mutually separable, further complicating calibration and signal extraction efforts.\\

Ongoing global $21$~cm experiments have primarily focused on frequencies ranging from $\sim100$~MHz to $\sim200$~MHz ($6<z<12$)\citep{bowmannature,chip09,saras,zebra} . Dark ages and cosmic dawn experiments at lower frequencies are being planned, or are being commissioned--- Dark Ages Radio Explorer (DARE) \citep{dare}, Large-Aperture Experiment to Detect the Dark Ages \citep{leda}, and Broadband Instrument for the Global HydrOgen ReionizatioN Signal (BIGHORNS) experiment \citep{bighorns}, among others. Global $21$~cm work especially in the lower frequency band requires a strict assessment of systematic chromatic corruptions. The reasons for this are threefold. (i) Firstly, the magnitude of ionospheric effects such as refraction and absorption increase rapidly with decreasing frequency. (ii) Secondly, Galactic foreground brightness temperatures increase with decreasing frequency as a power law with a relatively steep spectral index of about $-2.54$. Consequently, any systematic corruptions which are multiplicative, which many of the ones in such experiments are, may undermine $21$~cm signal detection efforts more severely at lower frequencies. (iii) Finally, the increased fractional bandwidth in the lower frequency band leads to an increased variation of antenna beams across the measurement bandwidth, giving larger chromatic effects.\\

Receiver gain and noise temperature may be calibrated by switching the receiver between the sky and known man-made noise sources. Such techniques have been demonstrated with moderate success in global $21$~cm experiments \citep{bowmannature,chip09,saras}. However, little attention has been paid in the literature to chromatic (frequency dependent) antenna beam and ionospheric effects. These effects have thus far been assumed to be `spectrally smooth' and possibly fitted away along with the foregrounds. They have thus escaped qualitative and quantitative treatment--- one of the primary aims of this paper.\\
% In this paper, we assume that the receiver gain and noise contribution have been calibrated, and evaluate the nature and extent of chromatic effects of the dipole beam, ionospheric refraction, and ionospheric absorption. We do so through realistic simulation of a single dipole experiment with a two layer ionospheric model corresponding to the F- and D-layers.\\

Chromatic effects must be studied in conjunction with algorithms that are used to separate the measured sky spectrum into foregrounds and the $21$~cm signal. Due to the lack of sufficiently accurate foreground models at these frequencies, such algorithms must rely on some priors on the differential properties of foregrounds and the $21$~cm signal. These algorithms typically exploit (i) the spectral smoothness of power-law-like foregrounds in comparison to less smooth structure expected in the $21$~cm signal, and/or (ii) the angular structure of foregrounds as opposed to isotropic nature of the global $21$~cm signal. Spectral smoothness of foregrounds may be exploited by casting the measured sky spectrum in a basis where foregrounds have a sparse representation unlike the $21$~cm signal. We may call such techniques spectral-basis methods, since they only use spectral smoothness as a prior. One such basis set suggested in literature, which we will call \texttt{logpolyfit}, uses polynomials in logarithmic space as basis to represent the time averaged spectrum \citep{bowman08,prit10,harker12}. Exploiting priors on the angular structure of foregrounds for global $21$~cm experiments has not received due attention in literature, save a recent effort by \citet{liu13}, who in light of their simulations, recommend measurements with an angular resolution of $\sim 5$~degrees. Practical implications of a narrow beam (highly chromatic sidelobes etc.) remain to be evaluated. Moreover, ongoing and proposed global $21$~cm experiments have near-hemisphere fields of view and lack any meaningful angular resolution. It is then instructive to place limits on the extent to which beam and ionospheric chromatic effects can confuse $21$~cm signatures in the context of spectral-basis algorithms--- another primary aim of this paper.\\

In this paper, we simulate the contribution of foregrounds (with chromatic effects) to the measured antenna temperature and evaluate an optimal set of basis functions for a sparse representation of foregrounds. By casting the foregrounds and the expected $21$~cm signal in this basis, we place limits on the amount of $21$~cm signal power that will be lost to foreground confusion in any spectral-basis technique. Since the optimal basis functions are not known apriori in real measurements, we may resort to predefined analytic basis functions such as polynomials. In this paper, we show that polynomials in logarithmic space (\texttt{logpolyfit}) are incapable of even detecting the presence of a template $21$~cm signal in simulated data in the frequency range of $45$~MHz to $80$~MHz.\\

In this paper, we propose an alternative spectral-basis method which we call \texttt{svdfit}, that evaluates a suitable basis using the measured data itself. Despite hemispherical fields of view of ongoing experiments, Earth rotation couples angular structure of the foregrounds into the time domain, while the global $21$~cm signal being isotropic, has no temporal structure. \texttt{Svdfit} uses the time variable component of the measured dynamic spectra to compute an efficient basis in which the foregrounds and chromatic effects have a sparse representation, but not the $21$~cm signal itself. We will show that a \texttt{svdfit} is better than \texttt{logpolyfit} in ascertaining the presence of a template $21$~cm signal in our simulated data. Nevertheless, we argue that for complete reconstruction of the $21$~cm signal spectrum, spectral smoothness is an inadequate prior in the above frequency range, and ultimately extracting the $21$~cm signal spectrum will require modeling of the foregrounds, antenna beam, and ionospheric effects via a full measurement equation.\\

The rest of the paper is organized as follows. Details of the simulations used herein are described in section \ref{sec:simu}. In section \ref{sec:iono} we describe our two-layered ionospheric model (F- and D-layers). We derive approximate expressions for chromatic effects from these two layers, and also quantify the level at which we expect these effects. In section \ref{sec:chrom} we use the results of our simulations to compute an optimal basis to represent the foreground induced antenna temperature, and quantify the extent to which foregrounds and chromatic effects confuse the $21$~cm signatures in spectral-basis methods. We then describe a novel foreground removal technique which we call \texttt{svdfit}, and also evaluate the efficiency with which \texttt{logpolyfit} and \texttt{svdfit} remove foregrounds and chromatic effects. Finally in section \ref{sec:concl} we draw conclusions and recommendations for future work.
%
%                             ___________
%                            /           \
% --------------------------- SIMULATIONS --------------------------
%                            \___________/
%

\section{Simulations}
\label{sec:simu}
This section describes the simulations used in this paper. We assume perfect bandpass calibration of receiver gain and receiver noise. We will discuss bandpass calibration in a forthcoming paper. We also assume that the antenna beam does not vary with time. This is a reasonable assumption since a dipole beam is a function of its mechanical shape, and hence, we do not expect noticeable variations in the antenna beam so long as the dielectric environment of the antenna does not change considerably. We build our simulations from smaller modules, each incorporating a different stage of signal corruption. The end result of our simulations is a dynamic spectrum measuring sky brightness temperature as a function of time and frequency.
% (see Figure \ref{fig:typical_spectrum}).\\
%
%                  /---------------\
% ----------------- MODEL PARAMETERS -------------------
%                  \---------------/
%
\subsection{Model parameters}
\begin{figure*}[h]
\centering
\includegraphics[width=\linewidth]{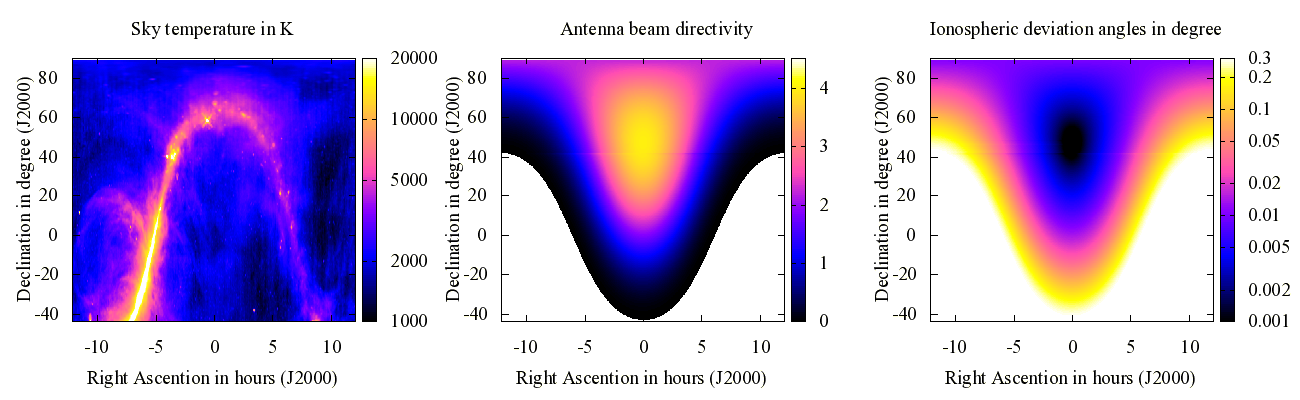}
\caption{Figure depicting the details of simulations presented in this paper at a frequency of $70$~MHz and LST $\approx 0$. Going from left to right, the images show, (i) the Haslam $408$~MHz all sky map scaled to $70$~MHz with a global temperature spectral index of $-2.54$, (ii) the simulated LOFAR LBA dipole beam, and (iii) the ionospheric refraction induced deviation angle for a homogeneous ionospheric model described in section \ref{sec:simu}\label{fig:simu}}
\end{figure*}

The following enumeration along with Figure \ref{fig:simu} describe the parameters in our modular simulations.
\begin{enumerate}
\item \emph{Location}: We assume the observation location to be one of the LOFAR stations (DE602) near Munich, Germany, for which we have data in hand \footnote{We have concluded a pilot study with data from DE602, and are currently acquiring science data}. The DE602 station is built on slightly sloping land. We assume the latitude and longitude of observations ($47^{\circ}47'9.77''$N, $9^{\circ}23'46''$E) to be that point on the locally flat ellipsoid which `sees' the same sky as DE602 at any instant of time. Though this corresponding point and DE602 have slightly different horizons, we discount this fact since we are primarily concerned with the chromatic effects of the beam and ionosphere in this paper.\\
\item \emph{Sky model}: Our simulations can use either of two sky models: (i) the $408$~MHz all sky map by \citet{haslam} scaled with a global temperature spectral index of $-2.54$, or (ii) the all-sky model by \citet{decosta} that is a linear combination of observations at different frequencies. While the second sky-model presents a more realistic scenario, we use the first sky model as a reference model to study the spectral nature of chromatic effects due to the ionosphere, LOFAR LBA dipole beam, and the foreground itself.\\ 
\item \emph{Antenna beam}: To study the effects of the chromatic LOFAR LBA dipole beam while facilitating comparison with previous work, we present simulations with two beams: (i) a frequency independent $\cos^2(\theta)$ beam ($\theta$ is the zenith angle) considered by \citet{prit10} in their simulations, and (ii) a realistic frequency dependent LOFAR LBA beam obtained from electromagnetic simulations\footnote{We used HFSS, a Finite Element Method (FEM) based full-wave 3D electromagnetic simulator.} of the antenna geometry including the finite ground-plane (see Figure \ref{fig:lba_pattern}). The frequency dependence of the antenna beam results in the sky being weighed differently at different frequencies and hence couples spatial structure on the sky into the frequency domain.\\
\begin{figure*}
\centering
\includegraphics{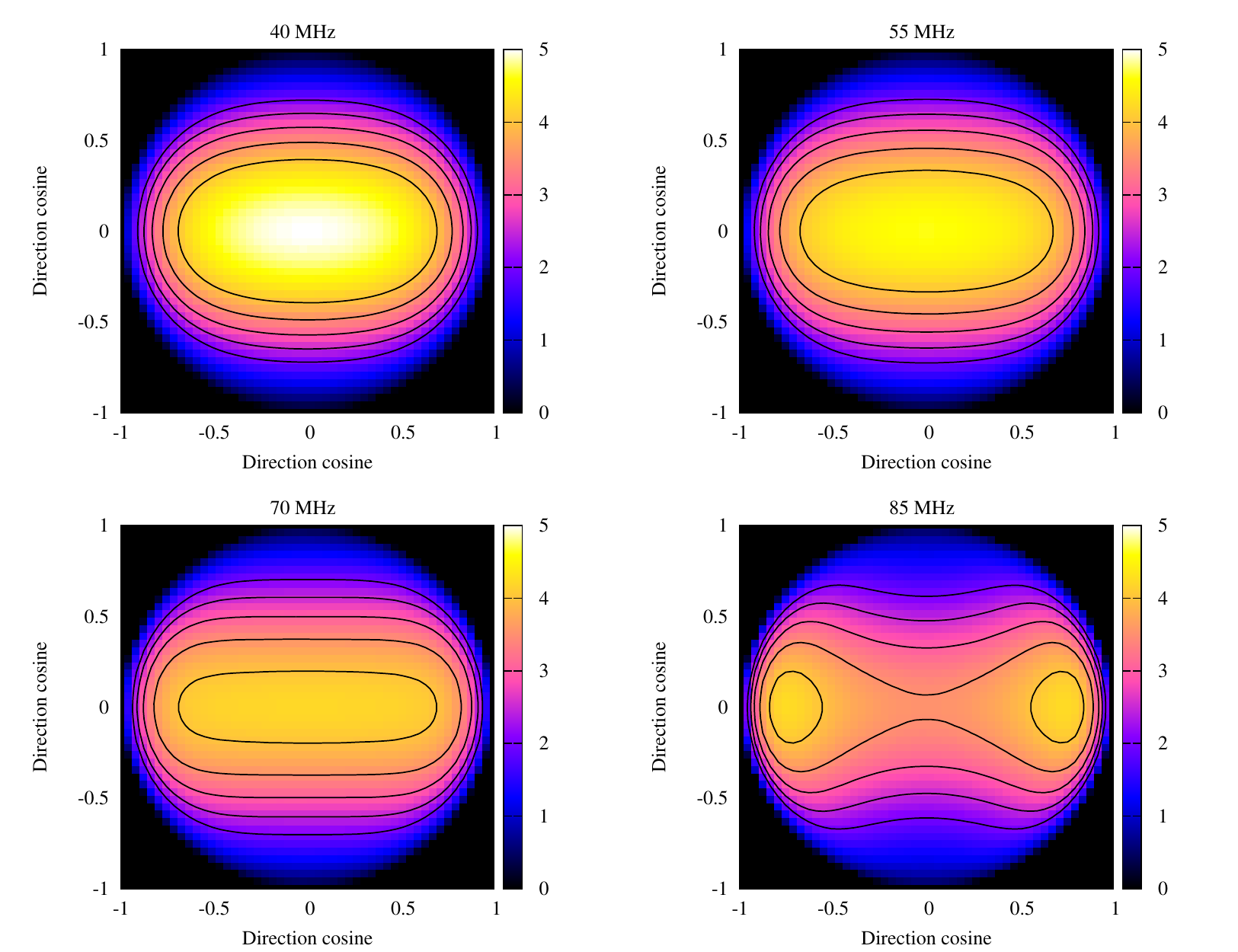}
\caption{Plots showing the variation of the simulated LOFAR LBA dipole beams with frequency. The $4$ panels show the directive-gain of the antenna at $4$ different frequencies: $40$~MHz (top-left), $55$~MHz (top-right), $70$~MHz (bottom-left), and $85$~MHz (bottom-right). Also overlaid are contours at directive-gain of $2$, $2.5$, $3$, $3.5$, and $4$ (outer to inner).\label{fig:lba_pattern}}
\end{figure*}
\item \emph{$21$~cm signal}: The main feature of the global $21$~cm signal expected in the $45$~MHz to $80$~MHz range is a relatively broad absorption feature \citep{furlanetto06a} \citep{prit08}. Since we do not address a full signal reconstruction here, we approximate this absorption feature as a negative Gaussian centered at about $70$~MHz, with FWHM of $7$~MHz. Spatial fluctuations of the signal are expected to be on small angular-scales ($<1$~degree), and are averaged away by the broad dipole beam. Recent work has shown that relative velocity between baryons and dark matter may imprint fluctuations on the observed $21$~cm brightness on $10-100$~Mpc scales \citep{bulkflows}, that correspond to several degrees in the sky. However, single dipoles typically have fields of view spanning several tens of degrees, and hence, we safely ignore any observable brightness fluctuations in the $21$~cm signal. \\

\item \emph{Ionospheric model}: Ionospheric effects may be divided into static effects and dynamic effects. Dynamic effects include time variant phenomenon such as (i) scintillation induced by turbulence in the ionospheric plasma \citep{crane77}, and (ii) refraction from large scale traveling ionospheric disturbances \citep{TIDs}. While we expect these effects to average away for long integration time scales, there has not been a comprehensive study thus far on their effects on global $21$~cm experiments, and we defer a discussion on these effects to a future paper. In this paper, we only study static ionospheric refraction and absorption. In section \ref{sec:iono}, we describe these static effects in detail. In particular, we show in section \ref{sec:iono} that a static ionosphere causes (i) frequency dependent deviation of incoming electromagnetic rays (chromatic refraction) which we model by `stretching' the antenna beam accordingly, and (ii) frequency and direction dependent absorption due to electron collision with air molecules which we model as a multiplicative loss factor on the antenna beam. These two effects, when applied on the fiducial antenna beam, give us an effective antenna beam which we use to compute the observed dynamic spectra. \\

\item \emph{Gridding and computation}: The sky temperature at a given time and frequency is computed by pixel wise multiplication of the sky model and the effective antenna beam, followed by integration of this product over all pixels while taking their solid angle into account. For a given epoch $t$, and frequency $\nu$, this computation may be represented as
\begin{equation}
\label{eqn:mmeq}
T_A(t,\nu) = \int_0^{2\pi}d\phi\int_0^{\pi/2}d\theta\sin\theta\, T_f(t,\nu,\theta,\phi)B(\nu,\theta,\phi),
\end{equation}
where $T_A$ is the simulated antenna temperature, $T_f$ is the sky brightness temperature (sky model) which is a function of zenith angle $\theta$ and azimuth angle $\phi$, and $B$ is the antenna beam as a function of frequency and sky position. Note that due to Earth rotation, $T_f$ changes with time. As mentioned before, ionospheric effects may be absorbed into the beam term, and if the beam in equation (\ref{eqn:mmeq}) is replaced by an effective beam $\hat{B}(\nu,\theta,\phi)$, then equation (\ref{eqn:mmeq}) is the measurement equation which describes the computations in our simulation.\\

Usually, the antenna beam and the sky model are specified in different co-ordinate systems, and have to be brought to a common grid to numerically compute a discretized form of equation (\ref{eqn:mmeq}). It is easier to re-grid and interpolate a smoothly varying antenna beam as opposed to the global sky model that has more complex structure due to the Galactic disk and point-like sources. We thus work in the co-ordinate system of the sky model (RA,DEC) and interpolate the effective antenna beams at each frequency and time epoch to the sky grid. An example sky model, antenna beam, and ionospheric refraction induced deviation angle on the sky grid is shown in Figure \ref{fig:simu} for a single frequency channel at a sidereal time of $00$h$00$m$00$s. For different values of sidereal time, the effective beam will be shifted along the Right Ascension axis. The end product of the simulations is a dynamic spectrum ($T_A(t,\nu)$ from equation \ref{eqn:mmeq}) in time-frequency domain. An example dynamic spectrum is shown in Figure \ref{fig:typical_spectrum}.
\end{enumerate}
\begin{figure}
\includegraphics[width=\linewidth]{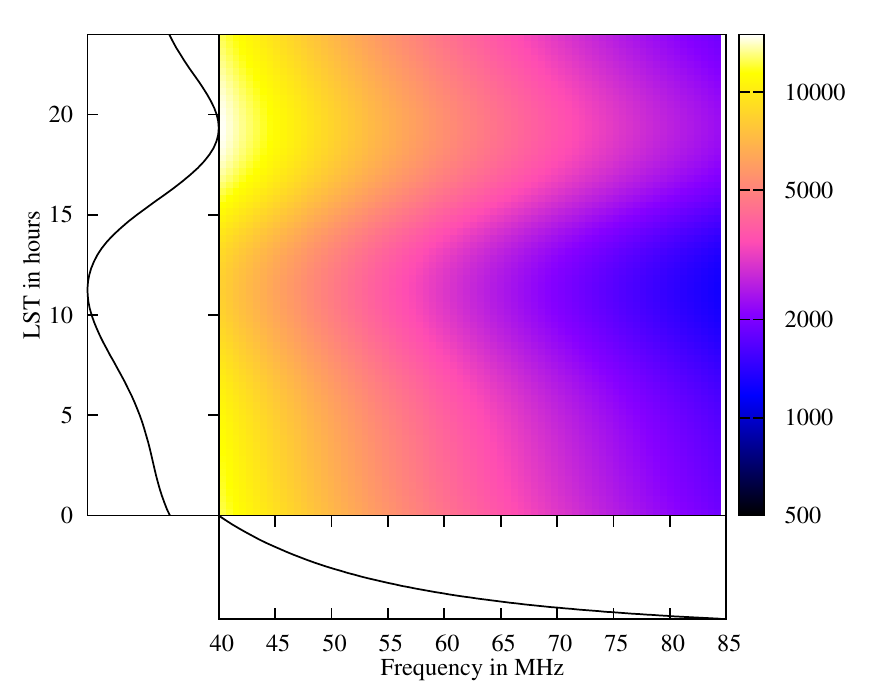}
\caption{Plot showing a typical dynamic spectra which is the output of our simulations. Color-bar units are in Kelvins of antenna temperature. Chromatic effects, and the $21$~cm  are too faint to be discerned by eye on this image. Also plotted on the left and bottom are averages along frequency and time axes respectively.\label{fig:typical_spectrum}}
\end{figure}
%
%                             ___________
%                            /           \
% ---------------------------THE IONOSPHERE --------------------------
%                            \___________/
%
%
\section{A static ionosphere}
\label{sec:iono}
This section describes the static model we use for the ionosphere in greater detail, and the origin and nature of chromatic refraction and absorption. The bulk of refraction and absorption occur at two separate layers of the ionosphere--- the F-layer and the D-layer respectively. We thus evaluate two refractive index values for typical conditions in these two layers, and then use the Earth-ionosphere geometry to compute the refractive ray deviation and the absorptive loss factor for different frequencies and directions.
%
%Refraction is a result of the dielectric properties of electron plasma, and absorption is a result of damping due to collisions between electrons and air molecules/ions. 
%
%(i) the F-layer, where most of the electron column density lies, and (ii) the D-layer where electron-gas molecule collision rate peaks. Due to its high electron density, the F-layer accounts for most of the chromatic refraction in the ionosphere, and due to its high electron collision rate, the D-layer accounts for most of the absorption in the ionosphere. 
%
%
\subsection{Ionospheric refractive index}
\label{subsec:refind}
The ionosphere is a magnetized plasma whose complex refractive index is given by the Appleton-Hartree equation\footnote{A German physicist by the name H. K. Lassen \citep{lassen} proposed a theory of propagation in a magnetized plasma before both Appleton and Hartree, but we use the name that is often found in literature.}  \citep{shkar61} that relates the refractive index to the electron density, magnetic field, and the geometry of wave propagation. We computed the change in refractive index due to the Earths magnetic field to be less than $1$ part in $10^4$ for the F-layer and less than $\sim 2$ percent for the D-layer. We ignore this effect since it is smaller than the refractive index variations induced by day to day changes in ionospheric electron density, and we present results for a broad range of electron densities. Additionally, the change in refractive index due to a magnetic field is different for left-hand and right-hand circularly polarized radiation, and results in an effect called Faraday rotation. In this paper, we ignore Faraday rotation by assuming the sky to be unpolarized on scales comparable to our antenna beam. We thus model the ionospheric refractive index using a simplified form of the Appleton-Hartree equation that does not include the magnetic field term:
\begin{equation}
\label{eqn:appl}
\eta^2 = 1-\frac{(\nu_p/\nu)^2}{1-\textrm{i}(\nu_c/\nu)},
\end{equation}
where $\nu_p$ is the electron plasma frequency, and $\nu_c$ is the electron collision frequency. The electric field of a plane wave traveling in a homogeneous ionospheric layer is given by 
\begin{equation}
E(\Delta s) \propto \exp\left(\frac{\textrm{i}2\pi\nu \Delta s}{c}\eta\right),
\end{equation}
 where $c$ is the speed of light in free-space, and $\Delta s$ is the distance measured along the direction of propagation in the ionosphere.\\

The real part of the refractive index $\eta$ is mostly sensitive to the electron density, and causes a change in the phase velocity (from that in free space) resulting in refraction. The imaginary part of $\eta$ (negative in our case) is mostly sensitive to the electron collision rate that in turn depends on the electron density, atmospheric gas density, and temperature. The imaginary part exponentially dampens the wave amplitude causing absorption.\\

Due to its low atmospheric gas density (giving a low collision rate) and high electron density, we model the F-layer with a real, frequency dependent refractive index $\eta_F$, to account for ionospheric refraction. Due to its high atmospheric gas density (giving a high collision rate) and low electron density, we model the D-layer with an imaginary, frequency dependent refractive index ($\textrm{i}\eta_D$) to account for ionospheric absorption.\\

We expect most of the global $21$~cm observations to take place during night-time when ionospheric electron density is at its lowest. Additionally, this avoids the need to model the complex and time-variant spectrum of the sun as a foreground source. We will thus assume values for a typical mid-latitude night-time ionosphere in the absence of intense solar activity. We refer the reader to \citet{thomps04}, \citet{evanhag}, and references therein from which we have drawn parameter values for typical ionospheric F- and D-layer conditions. The following subsections compute the refractive and absorptive effects of such an ionosphere.  
%
%
%                         /-------------------\
% ------------------------ F-LAYER REFRACTION ---------------------
%                         \-------------------/
%
%
\subsection{F-layer refraction}
\label{subsec:flayer}
Most of the ionospheric electron column density is accounted for by the F-layer that extends between a 
height of $\sim 200$~km and $\sim 400$~km from the Earth's surface. The electron density outside of 
this layer is known to fall off rapidly. Though the electron density does vary within the F-layer, to first order, we model the F-layer as a 
homogeneous shell between the heights of $200$~km and $400$~km. We assume a constant electron 
density of $5\times10^{11}$~m$^{-3}$ which gives a typical mid-latitude electron column density of 
$10$~TEC units\footnote{$1$ TEC units equals a column density of $10^{16}$ electrons per m$^2$. 
Ionospheric TEC is routinely monitored by measuring the propagation delay in GPS signals. 
See for instance http://iono.jpl.nasa.gov/latest\%5Frti\%5Fglobal.html}. This value is typical 
of winter-time in mid-latitudes where LOFAR is situated. TEC values are typically higher 
(i) during daytime, (ii) closer to the equator, and (iii) during summer. Additionally, ionospheric TEC is sensitive to solar and sunspot activity. Due to the above reasons, we will also 
present results for higher TEC values.\\

\begin{figure}
\centering
\includegraphics[width=\linewidth]{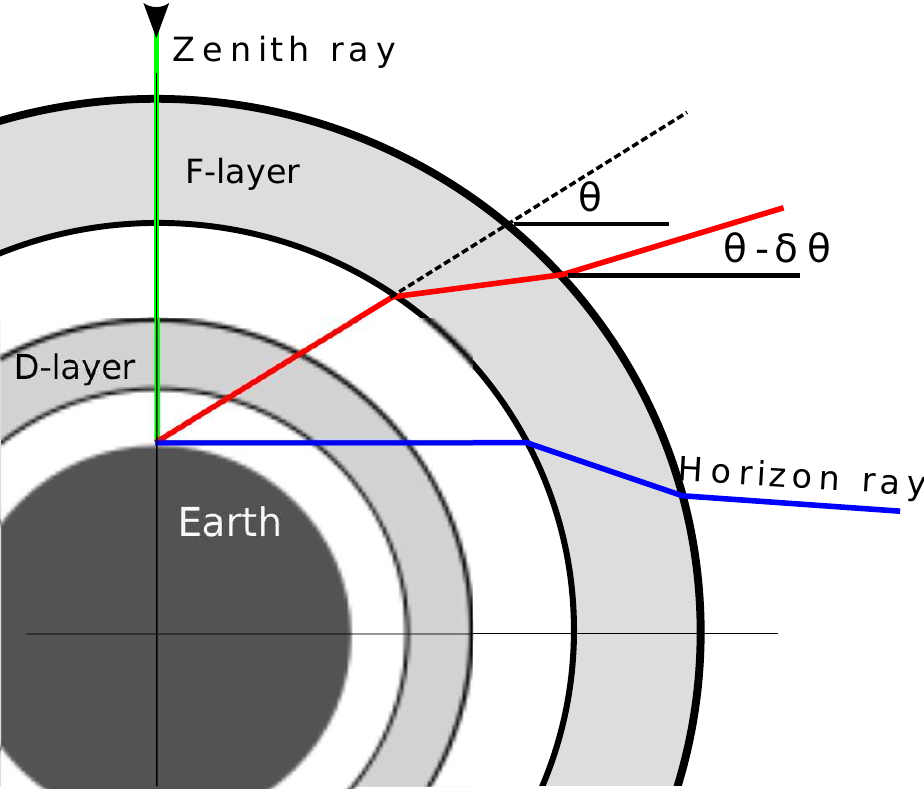}
\caption{Not-to-scale depiction of ionospheric refraction. The curvature of the Earth results in deviation in positions of sources in the sky. A homogeneous ionosphere thus acts as a lens \label{fig:refrac_illus}}
\end{figure}
Figure \ref{fig:refrac_illus} depicts the refractive effect of the ionospheric F-layer. Any incoming 
ray suffers Snell's refraction at the upper and lower boundaries of the F-layer. If the Earth were flat, there would be no net 
deviation in the ray. Due to the curvature of the Earth (and hence the ionosphere) there is a net 
deviation $\delta \theta$. This deviation is zero for a source at zenith, and increases as we move 
towards the horizon. Hence, the ionosphere acts like a spherical `lens' that deviates incoming rays towards zenith. Since the ionospheric refractive index is a strong function of frequency, 
$\delta \theta$ is also a function of frequency. Consequently, the ionosphere is a chromatic lens. 
Figure \ref{fig:refrac_illus} also depicts a horizon ray that marks the radio horizon, that is below the geometric horizon. 
This radio horizon is different at different frequencies. This chromatic lensing of the sky 
due to ionospheric refraction is an important effect for global $21$~cm experiments that 
use dipoles with near hemispherical fields of view.\\

It is difficult to derive a closed form expression for $\delta\theta(\nu,\theta)$, and hence we compute 
it numerically by applying Snell's law at the two interfaces. Nevertheless, we may use an 
analytical approximation to study the dependency of $\delta \theta(\nu,\theta)$, on $\theta$ 
and $\nu$ \citep{bailey48}:
\begin{equation}
\label{eqn:devangle}
\delta\theta(\nu,\theta) \propto \nu^{-2}\cos(\theta)\left(\sin^2\theta+\frac{2h_F}{R_e}\right)^{-1.5},
\end{equation}
where $h_F$ is the mean height of the F-layer ($300$~km in our case) and $R_e$ is the radius of the Earth which we assume to be $6300$~km.
\begin{figure}
\centering
\includegraphics{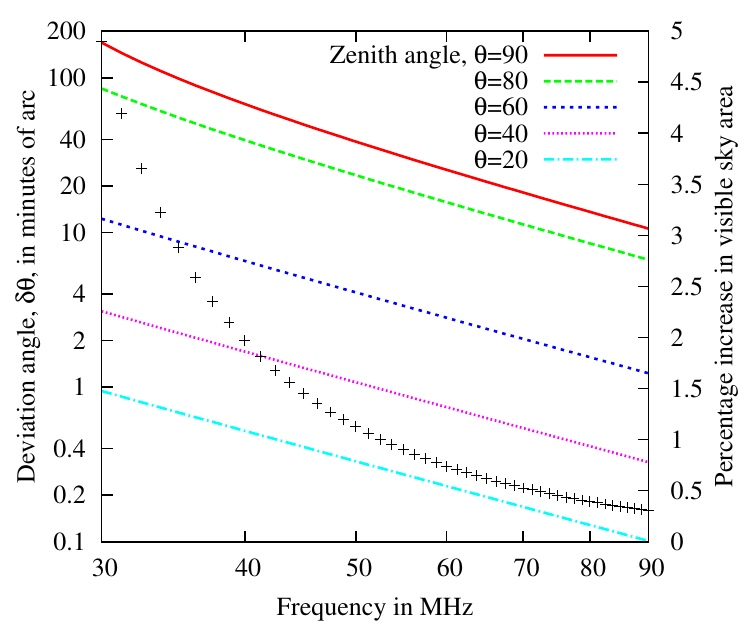}
\caption{Calculated deviation angle for a homogeneous ionospheric shell extending from  $R=200$~km 
to $R=400$~km with electron density $n_e=5\times10^{11}$m$^{-3}$. The deviation angle is a strong function 
of both incidence angle and frequency. Also shown in black `+` symbols is the percentage increase 
in sky area due to extension of radio horizon due to refraction.\label{fig:devangle}}
\end{figure}
Figure \ref{fig:devangle} shows a plot of $\delta \theta$ as a function of frequency for different 
elevation angles. These curves are obtained from simple ray tracing and not from equation (\ref{eqn:devangle}). The curves 
approximately follow a $\nu^{-2}$ dependence as expected. We also use 
the value of $\delta \theta$ for the horizon ray (see Figure \ref{fig:refrac_illus}) to compute 
the percentage increase in visible sky area as compared to the geometric horizon for each frequency. 
This percentage is also plotted in Figure \ref{fig:devangle}, and is not only frequency dependent, but also of the order of a 
few percent even for favorable ionospheric conditions. Since the foregrounds are $\sim 4-5$ orders 
of magnitude higher than the $21$~cm signal, a frequency dependent additional sky area of a few percent adds an amount of power 
which is $\sim 2-3$ orders of magnitude higher than the expected $21$~cm signal. Hence, it is 
important to consider the chromatic effects of ionospheric refraction\footnote{Note that the 
troposphere also causes refractive deviation by an angle $\sim0.35^{\circ}$ at the horizon and 
rapidly decreasing as we move towards zenith \citep{thomps04}. However this refraction is expected to be non-chromatic \citep{thomps04} and hence, we disregard it for the purposes of this paper.}.\\

The refractive lensing effect of the ionosphere may be absorbed into the antenna beam. We do this by `stretching' the antenna beam by an amount $\delta \theta$ to form an effective beam that now includes the effects of chromatic F-layer refraction. If the antenna beam is represented as $B(\nu,\theta,\phi)$, then the new effective antenna beam may be represented as
\begin{equation}
\label{eqn:effec_beam}
\hat{B}(\nu,\theta,\phi) =  B(\nu,\theta-\delta \theta,\phi),
\end{equation}
which gives the instantaneous measured antenna temperature of 
\begin{equation}
T_A(\nu) = \int_0^{2\pi}d\phi\int_0^{\pi/2}d\theta\sin\theta B(\nu,\theta-\delta\theta,\phi)T_f(\nu,\theta,\phi)
\label{eqn:anttemp}
\end{equation}
Note that the effective beam does not integrate to unity like the 
original antenna beam. It integrates to a value larger than unity due to the additional sky area added by refraction of sub-horizon rays into the original antenna beam. Since the dipole beam is `stretched' by an amount equal to $\delta \theta$ which is a strong function of frequency, the sky is weighted 
differently at different frequencies in equation (\ref{eqn:mmeq}). This couples spatial structure in the sky to frequency structure in the measured antenna temperature spectrum. It is important to note that this chromatic mixing happens even if the original antenna beam itself is frequency independent. \\

Finally, we quantify the approximate extent and nature of chromatic refraction by considering a simple case where (i) the sky brightness 
temperature is a power law with the same spectral index $\alpha$ everywhere ($T_f(\nu,\theta,\phi) \propto \nu^{-\alpha}$), and (ii) the original antenna beam is frequency independent and is given by 
$B(\nu,\theta,\phi) \equiv \cos^2(\theta)$. The effective antenna beam (due to chromatic refraction) is then given by 
\begin{equation}
\hat{B}(\nu,\theta) = \cos^2(\theta-\delta \theta),
\end{equation} 
which on Taylor-expansion about $\theta$ gives
\begin{equation}
\label{eqn:taylor_beam}
\hat{B} \approx \cos^2(\theta)+\delta \theta(\nu,\theta)\sin2\theta
\end{equation}
Note that the effective beam is now chromatic, while the original antenna beam is not.
Additionally, we have shown in equation (\ref{eqn:devangle}) that $\delta(\nu,\theta)$ has a form that is separable in $\nu$ and $\theta$, and may be expressed as 
$\delta \theta = \nu^{-2}g(\theta)$, where $g(\theta)$ is a function independent of $\nu$.
Substituting this and equation (\ref{eqn:taylor_beam}) into equation (\ref{eqn:anttemp}), we find that the antenna temperature evaluates to the form 
\begin{equation}
\label{eqn:f_final}
T_A(\nu) = F_1(\nu^{-\alpha} + F_2\nu^{-\alpha-2}),
\end{equation} 
where $F_1$ and $F_2$ are independent of frequency, and depend only on the sky brightness, antenna beam, and geometric terms. Equation (\ref{eqn:f_final}) shows that chromatic refraction 
will add a new component to the original $\nu^{-\alpha}$ sky brightness temperature. This new component has a spectral shape given by $\nu^{-\alpha-2}$, and as argued before, is at a few percent level. The chromatic foregrounds can now be fit away by the basis functions $\nu^{-\alpha}$ and $\nu^{\alpha-2}$. However, since the sky brightness and the LOFAR LBA beam are both more complicated, we will resort to more realistic simulations in section \ref{sec:chrom} to accurately evaluate the nature and extent of chromatic refraction.
%
%
%                      /------------------\
% --------------------- D-LAYER ABSORPTION ------------------------
%                      \------------------/
%
%
\subsection{D-layer absorption}
\label{subsec:dlayer}
The D-layer is the lowest layer of the ionosphere extending from $\sim 60$~km to $\sim 90$~km from 
the Earth's surface. High electron densities in the D-layer are expected to persist only during 
daytime due to solar insolation. However, residual electron densities of the order of
~$\sim 10^8$~m$^{-3}$ exists even during night-time. We will use a fiducial value for D-layer electron 
density of $5\times10^8$~m$^{-3}$ in our simulations. We choose this value to obtain an absorption of $0.01$~dB at $100$~MHz that agrees with values quoted in literature \citep{thomps04}. While this electron density is too low to cause appreciable refraction, due to a high atmospheric gas density at these heights, 
the electron collision frequency is high enough to cause considerable absorption. We use a typical value of 
$10$~MHz for the electron collisional frequency \citep{nico53}. We model the D-layer as 
a homogeneous layer between the heights of $60$~km and $90$~km.\\

\begin{figure}
\centering
\includegraphics{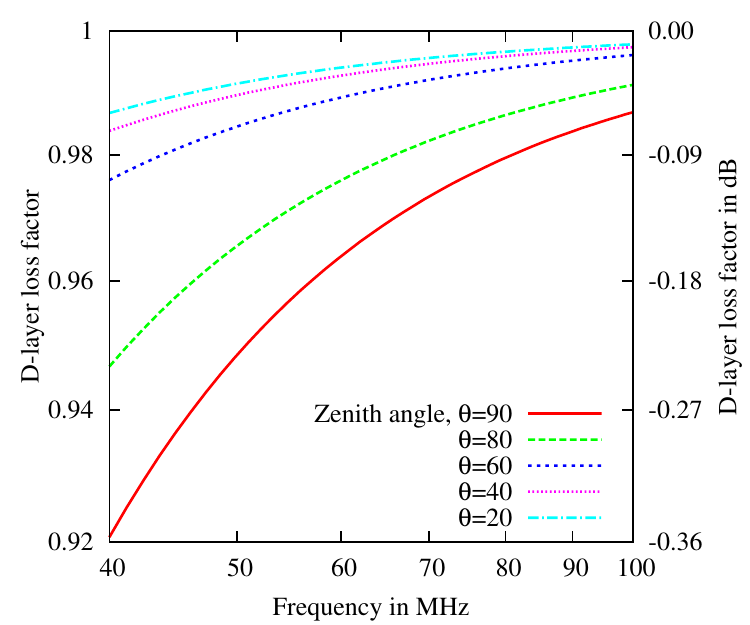}
\caption{Plot showing the multiplicative loss factor as a function of frequency due to absorption in the D-layer. We have assumed the D-layer to be a homogeneous shell with electron density of $5$x$10^8$m$^{-3}$, and electron collisional frequency of $10$~MHz, extending from a height of $60$~km to $90$~km.\label{fig:iono_abs}}
\end{figure}

If the path-length through the D-layer is $\Delta s$, then the multiplicative loss factor due to 
ionospheric absorption may be computed as 
\begin{equation}
\mathcal{L}(\nu,\theta) = \exp\left(2\pi\frac{\nu\eta_D}{c}\Delta s\right) \approx 1+2\pi\frac{\nu\eta_D}{c}\Delta s
\label{eqn:dlayer_abs}
\end{equation} 
where $\eta_D$ is the imaginary part of the D-layer refractive index, $c$ is the speed of light in vacuum, and the approximation holds for small values of the exponent (expected in the D-layer). Note that since $\eta_D$ is negative, $\mathcal{L}(\nu,\theta)<1$. We compute $\Delta s$ numerically in our simulations, but to understand its
dependencies, we provide an approximate expression:
\begin{equation}
\Delta s \approx \Delta h_D \left(1+\frac{h_D}{R_e}\right)\left(\cos^2(\theta)+\frac{2h}{R_e}\right)^{-\frac{1}{2}},
\label{eqn:path_length}
\end{equation}
where $R_e$ is the radius of the Earth, $h_D$ is the mean D-layer height ($75$~km in our case), and 
$\Delta h_D$ is the width of the D-layer. In Figure \ref{fig:iono_abs} we show the computed values of D-layer
absorption, $\mathcal{L}(\nu,\theta)$, as a function of frequency for different elevation angles. Zenith absorption increases from $\sim 0.01$~dB at $100$~MHz to $\sim 0.06$~dB at $40$~MHz.
As seen in the plot, absorption changes the incoming brightness temperature by $\sim 1-2$ percent. 
Though this is small, it is still large in comparison to the dynamic range between the foregrounds 
and the expected $21$~cm signal. Hence, studying the nature of D-layer absorption is important.\\

\begin{figure}
\includegraphics[width=\linewidth]{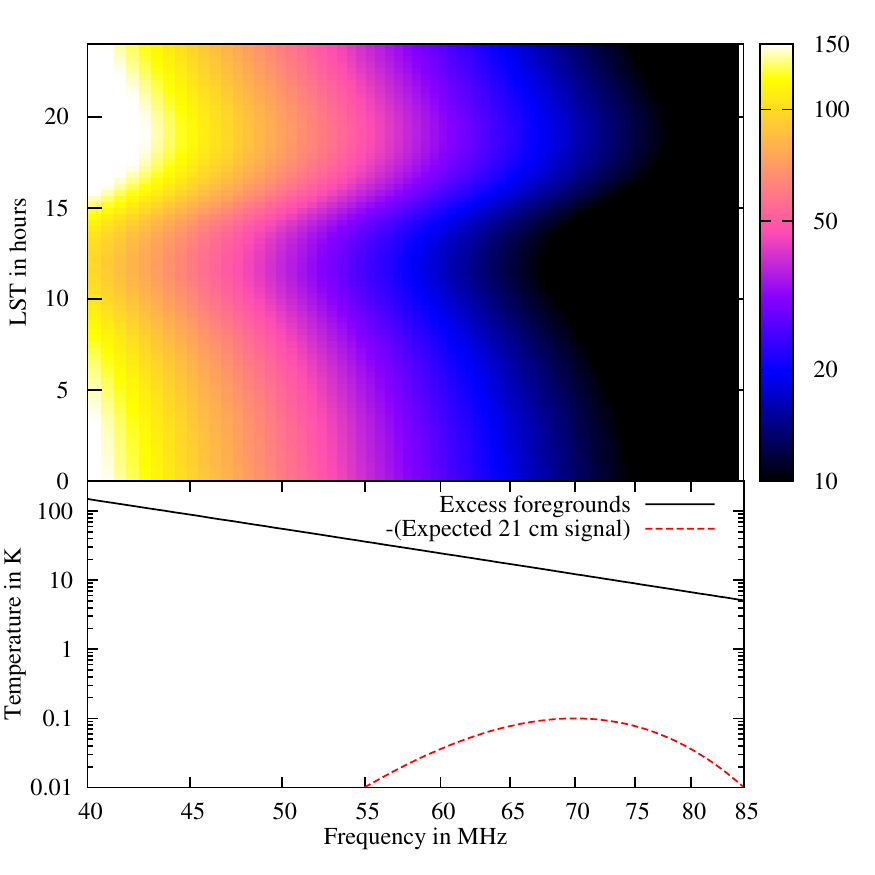}
\caption{Plot showing the differential antenna temperature due to the inclusion of ionospheric effects in our simulations. Top panel shows the difference between the simulated dynamic spectra obtained with and without the inclusion of ionospheric effects, and the bottom panel shows the differential spectrum averaged in time. Also shown for comparison is (the negative of) the expected $21$~cm signal. \label{fig:diff_spectrum}}
\end{figure}

The imaginary part of the refractive index from equation (\ref{eqn:appl}) approximately behaves as
\begin{equation}
\eta_D \approx -\frac{\nu_p^2\,\nu_c/\nu}{2(\nu_c^2+\nu^2)}.
\end{equation}
This along with equation (\ref{eqn:dlayer_abs}) gives
\begin{equation}
\label{eqn:dlayer}
\mathcal{L}(\nu,\theta) \approx 1-\frac{\pi \nu_p^2 \nu_c\Delta s(\theta)}{c(\nu_c^2+\nu^2)} 
\end{equation}
Since $\nu_c$ is comparable to frequency $\nu$, $\mathcal{L}(\mathcal{\nu,\theta})$ does not have a simple power-law like structure as a function of frequency (also evident in Figure \ref{fig:iono_abs}). As we did with refraction, we may define an effective beam $\hat{B}$ that takes into account 
ionospheric absorption. This beam will now integrate to less than one, and is given by
\begin{equation}
\hat{B}(\nu,\theta,\phi) = B(\nu,\theta,\phi)\mathcal{L}(\nu,\theta).
\end{equation}
For a frequency independent $\cos^2(\theta)$ beam, using equation (\ref{eqn:dlayer}), the effective beam is given by 
\begin{equation}
\hat{B}(\nu,\theta,\phi) = \cos^2(\theta)\left(1-\frac{\pi \nu_p^2 \nu_c\Delta s(\theta)/c}{\nu_c^2+\nu^2}\right).
\end{equation}
If the sky brightness temperature had a direction independent $\nu^{-\alpha}$- type power law behavior, the measured antenna temperature in presence of absorption will be of the form 
\begin{equation}
T_A(\nu) = D_1\left(\nu^{-\alpha}-D_2\frac{\nu^{-\alpha}}{(\nu_c^2+\nu^2)}\right),
\end{equation} 
where $D_1$ and $D_2$ are independent of frequency, and depend only on the sky brightness distribution, D-layer plasma and collision frequencies, and some geometric terms. This equation shows that, in case of D-layer absorption, we also have an additional component in the antenna temperature. This component has a spectral shape given by $\nu^{-\alpha}/(\nu_c^2+\nu^2)$, is negative, and as shown earlier, is at the $1-2$\% level. Because we expect $\nu_c$ to be $\sim 10$~MHz, discounting some error near the lowest end of our bandwidth, we may assume $\nu^2\gg\nu_c^2$, and the additional component may be approximated as having a spectral shape $\nu^{-\alpha-2}$, as in the case of F-layer refraction:
\begin{equation}
T_A(\nu) \approx D_1\left(\nu^{-\alpha}-D_2\nu^{-\alpha-2}\right),\,\,\, \nu\gg\nu_c.
\end{equation}
Consequently, for the case of a sky with a global spectral index measured with a frequency independent $\cos^2(\theta)$ beam, ionospheric effects (refraction plus absorption) will introduce an additional spectral contribution which approximately has a $\nu^{\alpha-2}$ shape. This is confirmed in Figure \ref{fig:diff_spectrum}, where we show the excess antenna temperature from our simulations due to the inclusion of ionospheric effects.
We have plotted the absolute value of the excess, since absorption over-compensates the additional sky coverage due to refraction, making the excess temperature negative. The additional antenna temperature due to F-layer refraction varies from $\sim 20$~K at $40$~MHz to $\sim 1$~K at $85$~MHz. The excess temperature due to D-layer absorption on the other hand varies from about $\sim -130$~K at $40$~MHz to $\sim -6$~K at $85$~MHz. As expected, the net differential temperature is at a $\sim 1$\% level. Also shown in the bottom panel is the time averaged excess temperature along-with (the negative of) the expected $21$~cm signal. Our simulations confirm that the chromatic effects due to a static ionosphere alone are $2$~orders of magnitude larger than the expected $21$~cm signal and hence must be studied with care. In Figure \ref{fig:diff_spectrum} we have neither included the chromatic effects of the LBA beam, nor a more comprehensive (and more complex) sky model by \citet{decosta}. Nevertheless, the figure stresses the fact that chromatic contamination that comes from just the ionosphere is larger than our signal by more than $2$ orders of magnitude. In section \ref{sec:chrom} we show results from simulations that also include the above two effects. 
%
%                          /---------------------\
% ------------------------- LOW ELEVATION CUT OFF --------------------------
%                          \---------------------/
%
\subsection{Low elevation reflection}
\label{subsec:leco}
\begin{figure}
\includegraphics[width=\linewidth]{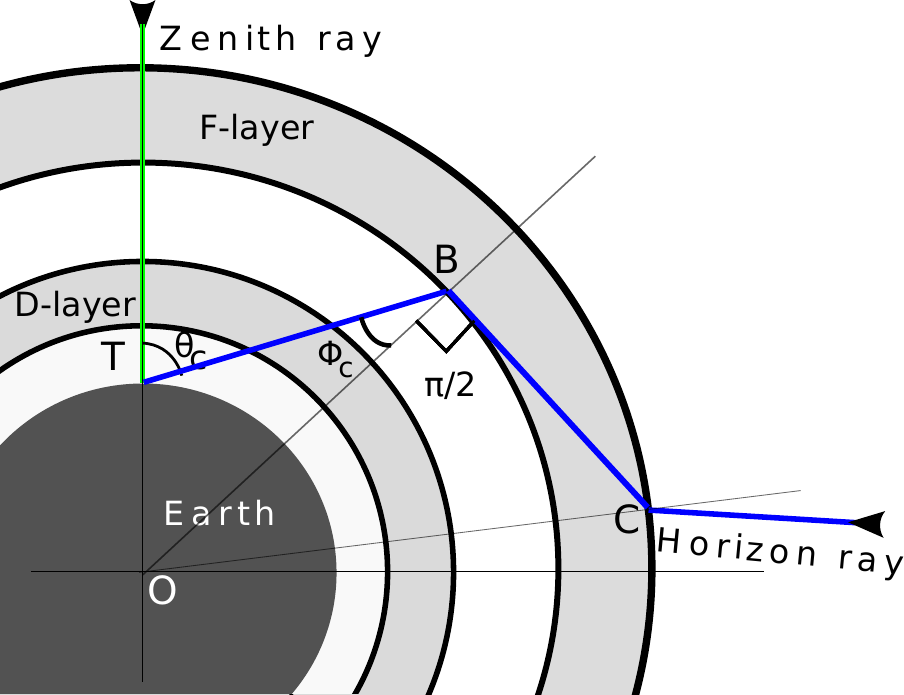}
\caption{Figure depicting the phenomenon of low elevation cutoff. The horizon ray reaches the lower F-layer interface at grazing incidence. Any incoming ray with a higher zenith angle is not incident at the lower F-layer interface at all, and simply escapes into space without ever reaching the telescope at point T.\label{fig:iono_tir}}
\end{figure}
Figure \ref{fig:iono_tir} depicts the effects of curvature of the ionosphere in cases when the F-layer 
electron density is considerably larger resulting in high deviation angles. We have thus far assumed that incoming rays that are refracted at the upper F-layer boundary are always incident on its lower boundary. Due to the Earth's curvature, this need not be true in cases when the F-layer electron density is particularly high. Consider the blue ray in Figure \ref{fig:iono_tir}. The ray is at grazing incidence at the lower F-layer interface marked as point B. Any ray that comes from space at a lower elevation angle will not be incident at the lower F-layer interface at all and will escape back into space without reaching the telescope at point T. Hence, in this case, the blue ray is `critical' and represents the horizon ray. From the point of view of the observer at T, all incoming rays have a zenith angle that is greater than some critical zenith angle $\theta_c \ge 0$. We will thus call this phenomenon low elevation cutoff\footnote{Conversely, any ray \emph{transmitted} from point T at zenith angle $>\theta_c$ will suffer total internal reflection at the F-layer--- an important consideration for communication links.} \\

We now compute the conditions under which a low elevation cut-off is relevant, and discuss its effect on the measured sky spectrum. For the case of the horizon ray, the angle of incidence at the lower F-layer interface is equal to $\pi/2$, and hence Snell's law at the interface becomes (see Figure \ref{fig:iono_tir})
\begin{equation}
\sin(\phi_c) = \eta_F
\end{equation} 
which upon using the sine rule in triangle OBT gives
\begin{equation}
\sin\theta_c = \frac{R_e+h}{R_e}\,\sin\phi_c = \frac{R_e+h}{R_e}\,\eta_F
\end{equation}
where $h$ is the height of the lower interface of the F-layer ($200$~km in our case). 
We may use the Appleton-Hartree formula (equation \ref{eqn:appl}) with $\nu_c=0$ in the above equation to get
\begin{equation}
\sin\theta_c = \frac{R_e+h}{R_e}\sqrt{1-(\nu_p/\nu)^2}
\end{equation}
Setting $\theta_c=\pi/2$ will then give us an expression for the frequency below which low elevation cut-off becomes relevant. Above this frequency, we may simply follow our earlier discussion of F-layer refraction from Section \ref{subsec:flayer}. Below this frequency, the increasing visible sky area due to increasing $\delta \theta(\nu,\theta)$ (chromatic refraction) is partly compensated by the decreasing sky area due to low elevation cut-off.  Setting $\theta_c=\pi/2$, gives 
\begin{equation}
\nu = \nu_p\,\frac{R_e+h}{h(2R_e+h)}.
\end{equation}
For our values of $h=200$~km and $R_e=6300$~km, this implies  $\nu\approx4\nu_p$.
Typical mid-latitude night-time F-layer plasma frequencies are below several MHz, and hence $\nu \approx 4\nu_p$ typically lies outside our bandwidth. In such cases, our discussion about the low elevation cut-off is not relevant. However, F-layer conditions where $\nu_p$ exceeds $10$~MHz do occur\footnote{Such conditions usually occur at low (Geo-magnetic) latitudes, and only during day-time at mid-latitudes. Global $\nu_p$ data may be found at http://www.ips.gov.au/HF\%5FSystems/6/5 and links therein.}, and unless due care is taken, some exposure to such conditions may persist in long integrations. \\

\begin{figure}
\includegraphics[width=\linewidth]{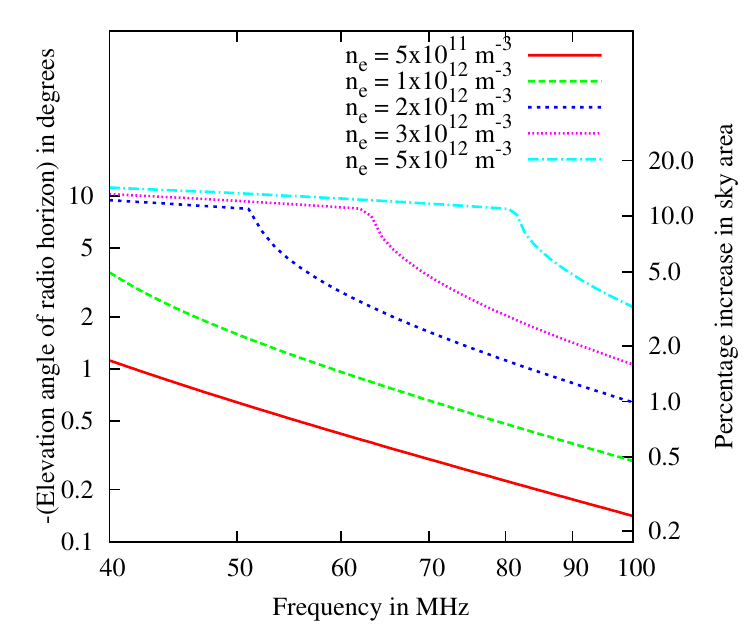}
\caption{Plot showing the elevation angle of the radio horizon versus frequency for various electron densities in the F-layer. The curves have a sharp knee at $\approx 4\nu_p$. For lower electron densities, the knee is below our minimum observation frequency, but not so for higher electron densities. Such an abrupt knee will introduce highly undesirable modulation in the foregrounds spectra.\label{fig:radio_horizon}}
\end{figure}

To illustrate the effect of low elevation cut-off, in Figure \ref{fig:radio_horizon}, we plot the elevation angle of the horizon ray as a function of frequency for different values of F-layer electron density. 
The electron density values for the different curves are $n_e=5\times10^{11}$, $1\times10^{12}$, $2\times10^{12}$, $3\times10^{12}$, and $5\times10^{12}$m$^{-3}$. These correspond to column densities of $10$, $20$, $40$, $60$, and $100$ TEC units respectively. The corresponding values of $4\nu_p$ are $25$, $36$, $51$, $62$, and $80$~MHz respectively. As expected, the first two curves have their $4\nu_p$ values outside of our bandwidth, and hence, do not suffer from the effects of low elevation cut-off within our bandwidth. The other curves, have a `knee' in their values of radio horizon positions versus frequency within the bandwidth. This knee exists because, as we go below the knee frequency ($\approx 4\nu_p$), the increasing visible sky area due to chromatic refraction is partly compensated by the decreasing visible sky area due to low elevation cut-off. Hence, the slope of the curves below and above the knee are markedly different. Such abrupt modulation of the foreground spectra is undesirable, given that the foreground subtraction algorithms rely on the smooth spectral nature of foregrounds at all times. It is also important to note that if such high electron density conditions persist even for a fraction of the total integration time, the time-averaged spectra will have components of foregrounds that are heavily modulated. It is thus important to monitor the ionospheric conditions throughout the observation duration, and perhaps even flag data acquired during times of high F-layer plasma density. 
%
%
%                           /--------------------------\
% ------------------------- EVALUATION OF CHROM EFFECTS -----------------------
%                           \--------------------------/
%
%
\section{Evaluation of chromatic effects}
\label{sec:chrom}
In section \ref{sec:iono}, we provided insight into the nature and extent of chromatic effects due to the ionosphere using simplistic dipole beams and sky models. Chromatic effects in presence of a more realistic sky model and dipole beam model are difficult to evaluate analytically. In this Section, we evaluate beam and ionospheric chromatic effects using the LOFAR LBA beam and more realistic sky models from \citet{decosta}. We will however also include results for a frequency independent beam for comparison with previous work, and for the benefit of experiments that use dipoles with (approximately) frequency independent beams (see \citet{saras} for instance).\\

How severe chromatic effects are depends on our prior knowledge of their nature and extent. If we have accurate enough models of foreground brightness and chromatic mixing, we may simply subtract the foreground contribution to antenna temperature (taking into account all chromatic effects) to expose the $21$~cm signal. This may not be the case in practice, and we are left to making certain simplifying assumptions about the differential properties of foregrounds (along with chromatic effects) and the $21$~cm signal. \\

The simplest assumption that we may make is that the foregrounds and chromatic effects have a smooth spectral behavior unlike the $21$~cm signal. We may then express the measured antenna temperature spectra in some optimal basis (polynomials for instance) wherein the foreground contaminants (but not the $21$~cm signal) are sparsely represented. We will refer to algorithms making only this assumption as spectral-basis methods. These methods assume no additional cognizance of the actual antenna beam, ionospheric effects, or constraints from other measurements such as interferometric visibilities.\\

For such spectral-basis methods, we will evaluate the chromatic effects at two levels. The first level inquires how similar the spectra of chromatic distortions are to the expected $21$~cm signal. The relevance of this question comes from the fact that chromatic effects are insidious if they confuse $21$~cm signatures in frequency space. To evaluate a best case scenario for spectral-basis methods, we will use the dynamic spectra from our simulations (excluding the $21$~cm signal contribution) to find an optimal basis to represent foreground contribution to the antenna temperature spectrum. We will then evaluate the relative efficiency with which the foreground contribution and the $21$~cm signal contribution to antenna temperature is fit away by these optimal basis vectors. This first level of inquiry merely evaluates the extent of foreground confusion. It does not give us a recipe to find an optimal basis in practice. At the end of this section we will briefly explore our second level of inquiry, that tries to evaluate two different spectral basis: (i) \texttt{logpolyfit} that uses polynomials in logarithmic space as basis function and (ii) \texttt{svdfit} (described later) which is a novel way to evaluate near-optimal basis functions from the dynamic spectrum itself.
%
%
%                         /----------------\
% ------------------------ AN OPTIMAL BASIS ------------------- 
%                         \----------------/
%
%
\subsection{An optimal basis}
\label{subsec:opt_basis}
\begin{figure*}
\includegraphics[width=\linewidth]{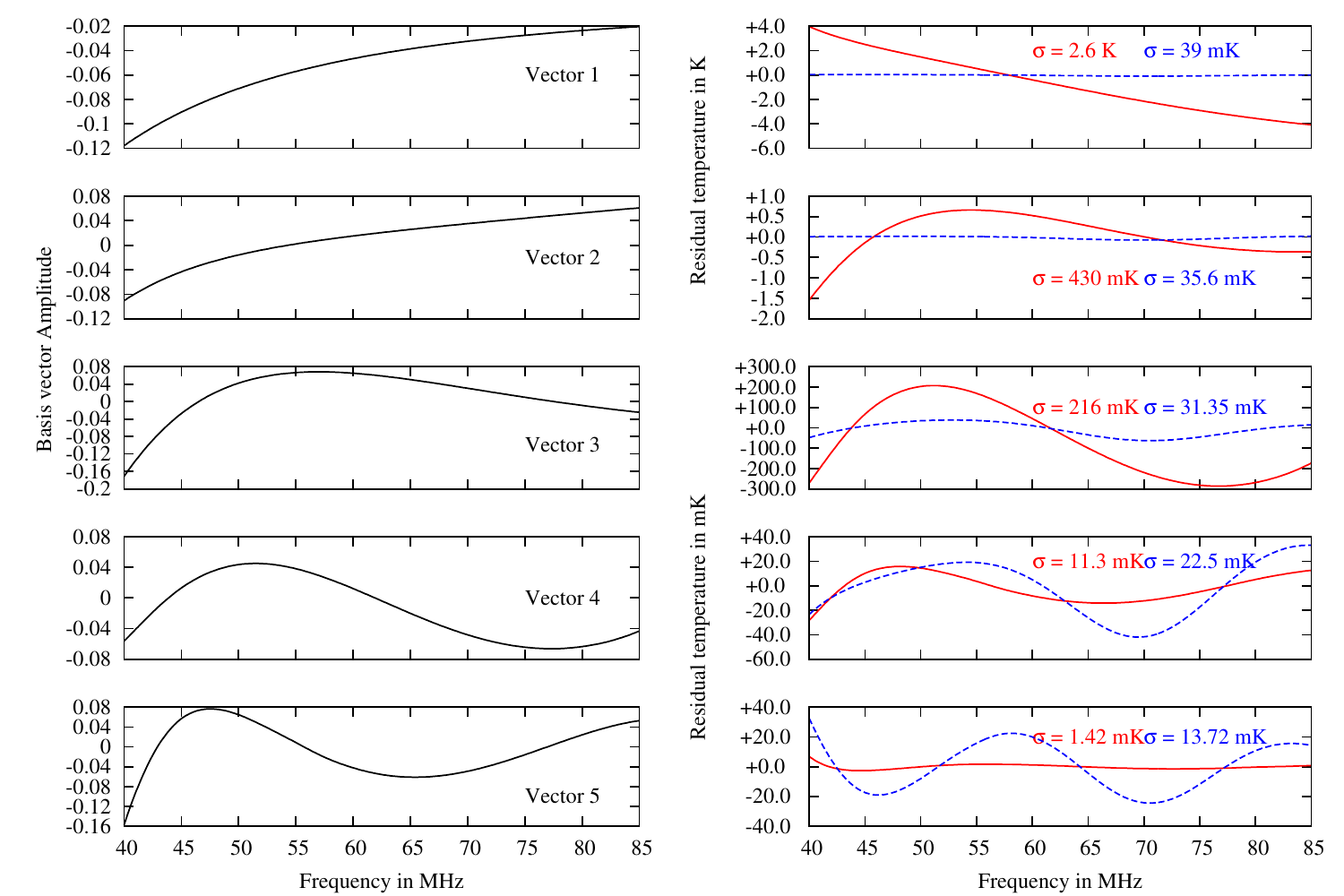}
\caption{Plot demonstrating the application of SVD in evaluating chromatic effects of the ionosphere. For this plot, the de Oliveira-Costa all sky map was used with the simulated LOFAR LBA beam. The left panels show the first $5$ dominant basis vectors ($\mathbf{v_1}$ through $\mathbf{v_5}$. Right panels show the residuals of representation of (i) the time averaged foregrounds (solid red) and (ii) the expected $21$~cm signal (broken blue) with the first 1,2,3,4, and 5 basis vectors.\label{fig:svd_demo_lba}}
\end{figure*}

%For instance, \citet{emma12} have used the Independent Component Analysis to compute efficient basis functions, albeit for the \emph{spatial} structure of foregrounds. \citet{liu12,liuteg12} have used the Singular Value Decomposition (SVD) to compute basis functions for the \emph{spectral} structure of foregrounds, albeit without chromatic effects. 

We compute an optimal non-parametric basis using the simulated antenna temperature without the inclusion of the $21$~cm signal. Note that in practice, with single dipole experiments, we do not have access to the antenna temperature without the inclusion of the $21$~cm signal, and hence we cannot compute an optimal basis from the data itself. \\

Non-parametric basis functions derived from the data itself have been explored in the literature, albeit in the context of interferometric observations of the spatial fluctuations of the $21$~cm signal. Examples include the Independent Component Analysis (ICA) \citep{emma12}, Singular Value Decomposition (SVD) \citep{liuteg12}, and smoothing techniques \citep{harker09}. More recently, \citep{liu13} have used the SVD technique in the context of global $21$~cm cm signal extraction. We will use the SVD technique to analyze chromatic effects in our simulations, due to its desirable orthogonality property.  \\

Our simulations provide a dynamic spectrum, $\mathbf{T_f}$, which is a matrix of dimensions $n_t$~x~$n_{\nu}$ where $n_t$ and $n_{\nu}$ are the number of time and frequency bins in the data respectively. We have used the subscript $\mathbf{f}$ to note that the dynamic spectrum used here is due to the foregrounds, and does not include the $21$~cm signal itself. We will use the subscript $\mathbf{21}$ to denote the measured spectrum due to the presence of the $21$~cm signal. To find an optimal basis where the foregrounds are sparsely represented, we compute the Singular Value Decomposition (SVD) of the dynamic spectrum $\mathbf{T_f}$:
\begin{equation} 
\mathbf{T_f} = \mathbf{U \Sigma V}^T.
\end{equation}
Matrix $\mathbf{V}^T = [\mathbf{v_1},\,\mathbf{v_2},\,...\mathbf{v_{n_{\mu}}}]$ is an orthonormal matrix of size $n_{\nu}$~x~$n_{\nu}$ whose rows ($\mathbf{v_i}$) provide an orthonormal basis to represent the spectral variability in $\mathbf{T_f}$. The vectors $\mathbf{v_i}$ are simply the eigen vectors of the correlation matrix $\mathbf{T_f}^H\mathbf{T_f}$. Hence, we are treating the spectra measured at different epochs as different realizations of snapshot measurements of sky spectrum, and through eigen decomposition, finding a set of basis vectors that efficiently describe any linear combination of these snapshot spectra. Since each snapshot spectrum has contributions from a large part of the sky, we expect the spectra to have some underlying ensemble properties that are efficiently described by the basis vectors $\mathbf{v_i}$.\\

The representation of the time averaged spectrum $\mathbf{t_f}$ in terms of a basis vector $\mathbf{v_i}$ is given by
\begin{equation}
\mathbf{t_f}(\mathbf{v_i}) = \mathbf{v_i}\mathbf{v_i}^T\mathbf{t_f},
\end{equation} 
and a representation of $\mathbf{t_f}$ in terms of the first $M$ basis vectors is given by
\begin{equation}
\mathbf{t_f}(\mathbf{V_M}^T) = \mathbf{V_M}\mathbf{V_M}^T\mathbf{t_f},
\end{equation}
where $\mathbf{V_M}^T = [\mathbf{v_1},\,\mathbf{v_2},\,...\mathbf{v_M}]$. Because the vectors $\mathbf{v_i}$ form an efficient basis to describe the foregrounds, we expect the residual rms given by $\textrm{rms}\left(\mathbf{t_f}(\mathbf{V_M}^T)-\mathbf{t_f}\right)$ to decrease rapidly as we increase $M$. Due to its contrasting spectral behavior, the time averaged $21$~cm spectrum, $\mathbf{t_{21}}$ is not expected to be efficiently represented by the above basis computed from the covariance matrix $\mathbf{T_f}^H\mathbf{T_f}$ due to the foregrounds alone. Consequently, we expect the residuals of its representation $\textrm{rms}(\mathbf{t_{21}}(\mathbf{V_M}^T)-\mathbf{t_{21}})$ to fall-off less rapidly with increasing $M$. The last two statements basically reiterate the sparseness assumption mentioned before. The value of $M$ for which the residual rms of foregrounds is lower than that of the $21$~cm signal then gives us the minimum number of parameters required to fit the foregrounds away. The difference in the rms of the $21$~cm signal and the rms of the $21$~cm residuals also gives us the amount of power in the $21$~cm signal that is fitted away along with the foregrounds. It might also come to pass that the rms of residual of foregrounds are always larger than that of the $21$~cm signal. In such cases, the foregrounds and chromatic effects are expected to introduce sufficient non-smooth structure, so as to inhibit their separation from the $21$~cm signal using only the information contained in the dynamic spectrum. In other words, the foreground spectral signatures will completely confuse our efforts to detect the $21$~cm signal with spectral basis methods. \\

Figure \ref{fig:svd_demo_lba} demonstrates the computation of basis functions and residuals using the SVD. For this figure, we have used the sky model from \citet{decosta}, the simulated LOFAR LBA beam, and $24$~hour observation over the frequency range $40$~MHz to $85$~MHz. We have included F-layer chromatic refraction ($n_e=5\times10^{11}$m$^{-3}$) and D-layer chromatic absorption ($n_e=5\times10^8$m$^{-3}$, $\nu_c=10$~MHz). The left-hand side panels show the first five basis vectors obtained from SVD. The basis vectors are arranged from top to bottom in decreasing order of dominance. The right hand panel shows the residuals when the time averaged antenna temperature spectrum due to foregrounds alone (red) and due to the expected $21$~cm signal (blue) are represented in terms of the first $1,2,3,4$ and $5$ dominant basis vectors (from top to bottom). The sky model from \citet{decosta} at our frequencies of interest is constructed from three principal components, and as such, may be fully expressed as a linear combination of three spectral basis vectors. The beam and ionospheric chromatic effects however, add additional complexity to the time-averaged antenna temperature spectrum and as a consequence, need at least $5$ basis vectors to be described to the required level. This is clear from the right-hand panel plots in Figure \ref{fig:svd_demo_lba} and the enumerated rms residual levels for the red and blue curves. Clearly, only for the case of representation with $5$ basis vectors does the rms residuals for foregrounds reduce to levels significantly below those of the $21$~cm signal. The expected $21$~cm signal has a residual of $\sim 13.7$~mK after fitting with $5$ basis vectors. The original $21$~cm signal that was used in the simulations had an rms (mean subtracted) of $34.85$~mK. This means that if we use spectral-basis methods, then not only we need a minimum of $5$ components to fit the foregrounds, we also end up loosing at least $85$\% of the variance (not rms) in the $21$~cm signal to foreground subtraction. 
%
%
%                     /------------------------------\
% ------------------- LIMITS OF SPECTRAL-BASIS METHODS -----------------
%                     \------------------------------/
%
%
\subsection{Limits of spectral-basis methods}
\label{subsec:limits}
\begin{figure*}
\includegraphics[width=\linewidth]{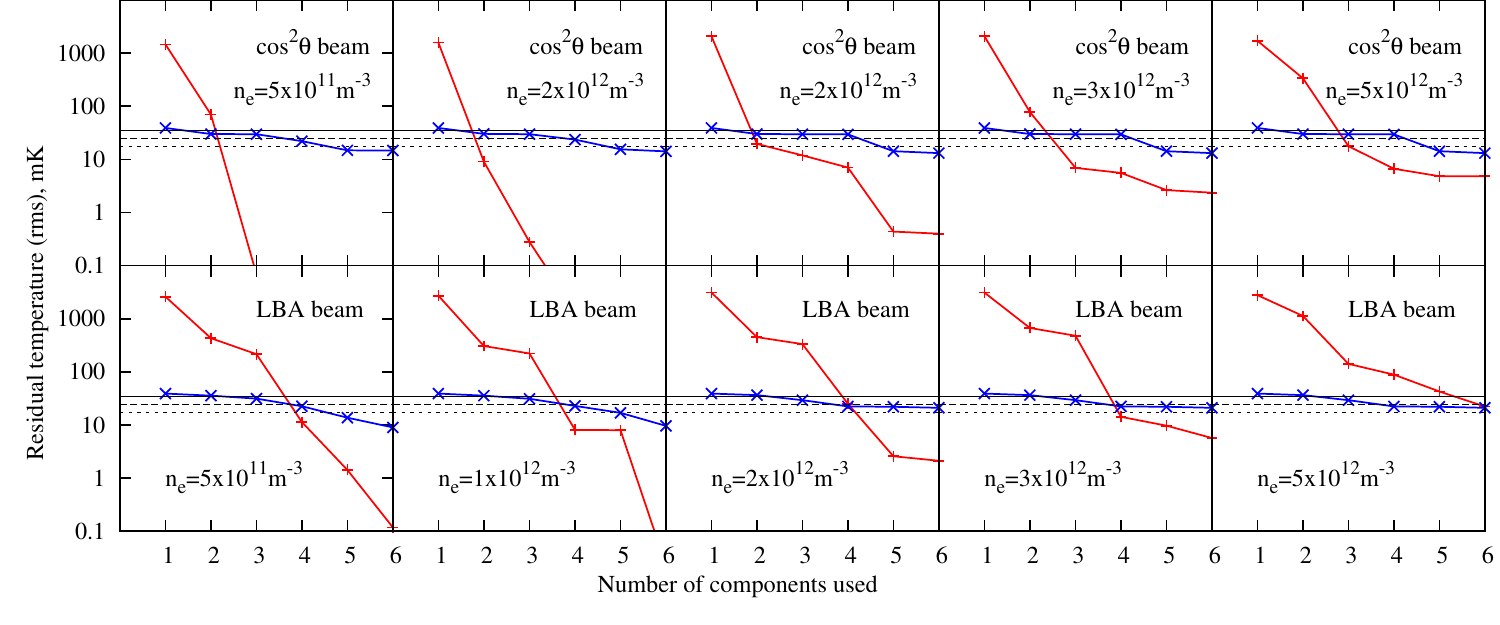}
\caption{Residual rms temperature when the mean antenna spectrum due to foregrounds (red `+'marks) and due to the fiducial $21$~cm signal (blue `$\times$' marks) are fit with an increasing number of basis vectors. All panels contain data simulated using the \citet{decosta} sky models. The top panels represent data for increasing values of F-layer electron density for a frequency independent $\cos^2(\theta)$ beam, and the bottom panels are for the simulated (frequency dependent) LOFAR LBA beams. The sold black line marks the $100$\% of the variance in the expected $21$~cm signal, and the broken and dotted black lines show levels corresponding to $50$\% and $25$\% of the original variance respectively.\label{fig:svd_rms}}
\end{figure*}

We now repeat the analysis of Section \ref{subsec:opt_basis} for the two dipole beams (i.e. achromatic and chromatic) and for different F-layer electron densities. The results are shown in Figure \ref{fig:svd_rms}. For brevity, we have not shown the basis vectors and residual spectra, but rather shown the rms of  residuals when (i) the foreground contribution to time-averaged antenna temperature (red) and (ii) the $21$~cm contribution to antenna temperature (blue) are fit with increasing number of spectral basis vectors. The top panels show the rms residuals for a frequency independent $\cos^2(\theta)$ beam, for different values of F-layer electron density. The bottom panels show the same for the fiducial LOFAR LBA beam obtained from electromagnetic simulations. As argued before, the abscissa for which the red curve dips below the blue curve denotes the minimum number of spectral basis vectors required to separate the foregrounds from the $21$~cm signal. Note however that this does not necessarily mean that the residuals are a representation of the $21$~cm signal as is evident from Figure \ref{fig:svd_demo_lba}. Each panel also show the levels denoting the $100$\% (solid black), $50$\%, and $25$\% (broken black) of the variance in the original $21$~cm signal spectrum (black line) to give intuition into the amount of variance (not rms) in the $21$~cm spectrum that is lost to foreground subtraction. Figure \ref{fig:svd_rms} thus also quantifies the minimum extent to which chromatic effects confuse the $21$~cm signature in spectral basis methods. Non-optimal basis functions (including analytic functions) not only require more parameters to model the foregrounds, but will also result in higher co-subtraction of the $21$~cm signal along with the foregrounds. \\

As mentioned before, in the absence of chromatic effects, the \citet{decosta} models may be fully described by just the first $3$ spectral basis vectors. Figure \ref{fig:svd_rms} shows that for a frequency independent beam, if the F-layer electron density is low ($<2\times10^{12}$m$^{-3}$), then the first $3$ spectral basis vectors are still sufficient to separate foregrounds and chromatic effects from the $21$~cm signal. In case of the LOFAR LBA beam, at least $4$ spectral basis vectors are necessary. Any generic basis functions (polynomials for instance) if employed, will only require more than the minimum number of parameters thus obtained from Figure \ref{fig:svd_rms}. For instance, for the case of the LOFAR LBA beam with $n_e=5\times10^{-11}$m$^{-3}$ (bottom-left panel), a minimum of $4$ parameters, or at least a $3$rd order polynomial is required. Indeed, there is no guarantee that a polynomial basis will represent the foregrounds sufficiently, but Figure \ref{fig:svd_rms} gives us a hard lower limit on the number of independent parameters required for foreground subtraction, and also the minimum amount of $21$~cm signal variance that will be lost to foreground subtraction with a spectral-basis method.
%
%
%                /---------------------\
% --------------- LOGPOLYFIT AND SVDFIT -------------------
%                \---------------------/
%
%
\subsection{\texttt{logpolyfit} and \texttt{svdfit}}
\label{subsec:fits}
\begin{figure}
\includegraphics[width=\linewidth]{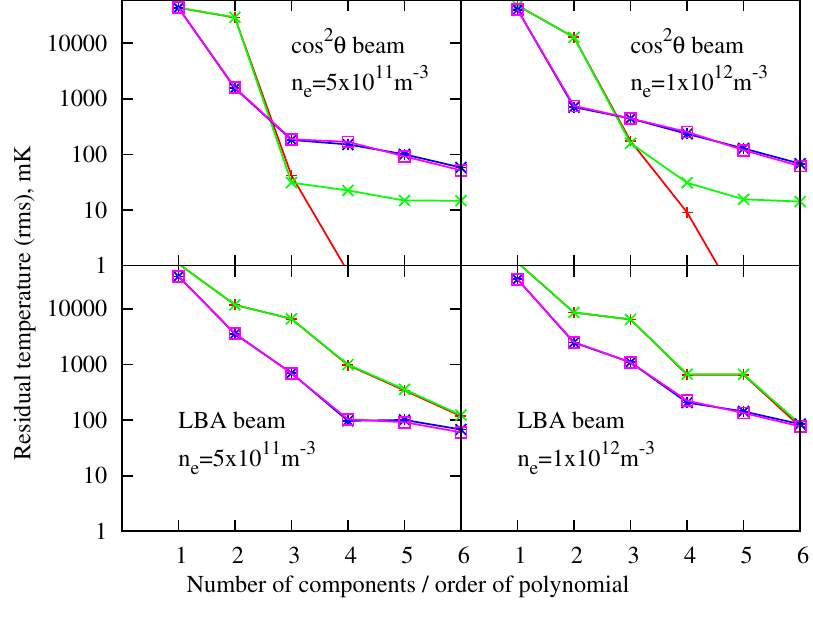}
\caption{Figure showing the rms residual of fit for the cases of \texttt{logpolyfit} (blue curve: Model 1, magenta curve: Model 2), and \texttt{svdfit} (red curve: Model 1, green curve: Model 2). While \texttt{logpolyfit} fails to separate the two Models in all $4$ panels, \texttt{svdfit} succeeds in the case of a frequency independent beam. \label{fig:fit2mthd}}
\end{figure}
The analysis of Section \ref{subsec:limits} presented an estimate of how much foregrounds and chromatic effects confuse $21$~cm signatures for spectral-basis methods. In reality, for single dipole experiments, we do not have access to the optimal basis functions simply because, we cannot obtain a measurement of antenna temperature that includes foreground contribution but not the global $21$~cm signal contribution. If the dipole is embedded into an array, then we will have independent measurements of spatially varying foregrounds. For single dipole experiments however, we have to adopt other techniques to construct the spectral basis vectors. In this subsection, we explore two such techniques. We will call the first technique \texttt{logpolyfit}. This method has been proposed in literature \citep{bowman08,prit08,harker12}, and attempts to use polynomials in logarithmic space as basis vectors to represent foregrounds. In other words, we seek to fit the time-averaged antenna temperature due to the foregrounds with the function
\begin{equation}
\mathcal{M}(\nu,a_0,a_1,a_2,...) = 10^{a_0(\log\nu)^0+a_1(\log\nu)^1+a_2(\log\nu)^2...}
\end{equation}
where $\log$ denotes logarithm to base $10$, and $a_0,a_1,a_2...$ are parameter values to be estimated. To evaluate (in simplistic terms) how well \texttt{logpolyfit} performs, we will fit the model $\mathcal{M}$ to two time-averaged antenna temperature spectra which we obtain from our simulations: (i) $\mathbf{t_{A}}(\nu)-\mathbf{t_{21}}(\nu)$, and (ii) $\mathbf{t_A}(\nu)$. Here $\mathbf{t_{21}}$ is the antenna temperature due to the fiducial $21$~cm signal, and $\mathbf{t_A}$ is the total antenna temperature due to both foregrounds and the $21$~cm signal. Hence the two models may be represented as
\begin{equation}
\mathbf{t_{A}}-\mathbf{t_{21}} = \mathcal{M}(\nu,a_0,a_1...) \,\,\, \textrm{: Model 1}
\end{equation}
and 
\begin{equation}
\mathbf{t_A} = \mathcal{M}(\nu,a_0,a_1,...)\,\,\, \textrm{: Model 2}
\end{equation}
Assuming that $\mathcal{M}(\nu,a_0,a_1...)$ fits out the foregrounds but not the $21$~cm signal, Model 1 must perform significantly better compared to Model 2 in terms of goodness of fit. This is because, Model 1 acknowledges the presence of the $21$~cm signal contribution in $\mathbf{t_A}$, whereas Model 2 does not. Note that we are not trying to fit for or solve for the $21$~cm signal spectrum. We are assuming its form and are merely trying to ascertain its presence in the simulated antenna temperature spectrum. In reality, we have to estimate the shape of the $21$~cm spectrum. But the above goodness of fit test provides a necessary condition for any spectral-basis algorithm to work in the first place. We evaluate the goodness of fit by computing the rms of the residuals of fit. \\

The second technique which we will call \texttt{svdfit} is a novel technique which we propose herein. \texttt{svdfit} derives basis vectors to describe the antenna temperature spectrum using the data itself (output of the simulation in our case). Since as argued before, the data contain the $21$~cm signal, evaluating the basis vectors from the data itself will give us a set of basis vectors that will remove the foregrounds and the $21$~cm signal with equal efficiency, thereby impeding their separation. We instead subtract the time averaged spectrum from the dynamic spectra prior to SVD evaluation. Since the antenna temperature due to a global $21$~cm signal is unchanging with time, the new mean subtracted dynamic spectra has no contribution from the $21$~cm signal, and we may compute a set of optimal basis vectors that will not efficiently describe the $21$~cm signal. We may then use these basis vectors to fit Model 1 and Model 2.\\

Subtracting the time-averaged antenna temperature from the dynamic spectra also removes any contribution from the global component of the foregrounds. In essence, we are extracting a set of basis from the spatially structured foregrounds, and using them to fit the away the global foreground component. Since the basis vectors are evaluated from only a fraction of the sky brightness temperature, they will be sub-optimal. However, our antenna beam averages over large parts of the sky, and since the sky brightness temperature spectrum is more or less stationary on such large scales, we expect these basis vectors to well represent the foreground antenna temperature contribution from the entire sky. \\

Figure \ref{fig:fit2mthd} shows the rms residuals for Model 1 and Model 2 for \texttt{logpolyfit} and \texttt{svdfit}. On the x-axis is the number of basis vectors used in case of \texttt{svdfit}, or the order of the polynomial used in case of \texttt{logpolyfit}. Model 1 and Model 2 rms for \texttt{svdfit} are plotted in red and green respectively, whereas, Model 1 and Model 2 rms for \texttt{logpolyfit} are plotted in blue and magenta respectively. The different panels in the Figure are for different F-layer electron densities, and the two different beams mentioned before. A successful algorithm must show a clear difference in the goodness of fit (represented by rms residuals here) for the two models. Note that in addition, the difference in rms between the Models must be larger than the uncertainties due to noise, which are not considered here. In any case, \texttt{logpolyfit} (blue and magenta curves) is not able to separate the two Models for any of the cases, and is hence an inadequate basis. \texttt{svdfit} on the other hand is able to ascertain the presence of the $21$~cm signal for the case of an ideal frequency independent beam. It however fails for the case of the fiducial LOFAR LBA beam. This is perhaps because of the highly chromatic nature of the LOFAR LBA beam which mixes spatial structure of the foregrounds into spectral structure in the antenna temperature spectrum. Consequently, basis vectors evaluated from the spatially structured foregrounds are rendered inefficient in fitting the spectrum due to the global foreground component.  Hence, \texttt{svdfit} may be an interesting technique only for experiments in our frequency range that use a (more or less) frequency independent antenna.
%
%              /-----------\
% ------------- WAY FORWARD -----------------
%              \-----------/
%
\subsection{Way forward}
\label{subsec:theway}
The simulations presented in this paper make simplistic assumptions about a more complicated reality. In particular, the real sky brightness may have much more complex spectrum than the one we have assumed in our simulation. Similar is the case for the ionospheric effects which in reality are dynamic and originate from an ionosphere structure that is more complicated than our simple layered model. The results presented herein are therefore expected to be lower limits to the chromatic mixing due to the foregrounds, beam, and the ionosphere. Consequently, spectral-basis methods may not be the proper way forward for foreground subtraction in our frequency range, despite being attractive due to the simplicity of their only assumption--- that the foregrounds have a smooth spectrum compared to the $21$~cm signal. The way forward warrants for techniques that use stronger priors on the properties of foregrounds and chromatic effects. \citet{liu13} have reached a similar conclusion in their analysis, and have suggested that spatial structure in the foreground be exploited to separate them from the $21$~cm signal that has no spatial features. This however requires higher spatial resolution unafforded by single dipole experiments. Antennas or arrays that have higher resolution also have highly chromatic beams with chromatic sidelobes that will cause a high level of chromatic mixing on many spectral scales. Algorithms that overcome this limitation have not been demonstrated till date.\\

Experiments with wide field-of-view dipoles may still benefit considerably from using stronger priors. This is especially true since we have explicit prior knowledge of (i) the sky brightness temperature from various surveys, (ii) fairly accurate antenna beam models from simulations and measurements, and (iii) knowledge of ionospheric conditions from GPS satellite based measurements or dipole-dipole cross-correlations. The LOFAR LBA dipoles are part of an array and inter-dipole visibilities may provide strong priors on all three of the above parameters. We thus propose to improve on spectral-basis methods in our future work, wherein we plan to model the data using a full measurement equation similar to equation (\ref{eqn:mmeq}).

\section{Conclusions and future work}
\label{sec:concl}
Single antenna wide field of view experiments that measure the sky spectrum to high degree of accuracy are interesting probes of the cosmic dark ages, cosmic dawn, and the epoch of reionization. Such measurements not only require unprecedented accuracy in receiver gain and noise temperature calibration, but also require accurate modeling and removal of contamination along the frequency axis. Ionospheric refraction and absorption are highly frequency dependent and are thus a potential limitation in such experiments. Additionally, a frequency dependent antenna beam mixes spatial structure of foregrounds into spectral structure providing an additional source of signal contamination. In this paper, we have studied the nature and magnitude of the above spectral contaminants in the frequency range of $40$~MHz to $100$~MHz that is particularly interesting for cosmic dawn studies. \\

We have arrived at the following results/conclusions:
\begin{enumerate}
\item A simple ionospheric model that accounts for static chromatic effects consists of two homogeneous layers: the F-layer that accounts for chromatic refraction, and the D-layer that accounts for chromatic absorption. In case of a sky with a global spectral index $-\alpha$, and a frequency invariant beam, chromatic refraction due to typical F-layer electron densities ($n_e=5$x$10^{11}$m$^{-3}$) adds a component to the measured antenna dynamic spectrum that has a spectral index of $-\alpha-2$, and is about $\sim 20$~K at $45$~MHz and $\sim 1$~K at $85$~MHz. Likewise, typical D-layer absorption for an electron density of $n_e=5\times10^8$m$^{-3}$, and collisional rate of $\sim10$~MHz, adds a (negative) component to the dynamic spectrum at the $\sim 1-2$\% level with spectral shape given by $\nu^{-\alpha}/(\nu_c^2+\nu^2)$. Typical values for additional signal due to D-layer absorption range from $\sim -130$~K at $40$~MHz to $-6$~K at $85$~MHz. We have also shown that high F-layer electron densities lead to low elevation cut-off of incoming rays, and that this leads to a knee-like feature in elevation angle of the horizon ray versus frequency. We have identified the knee to be at $\sim 4$ times the plasma frequency. We have shown that if the F-layer electron densities approach or exceed $n_e=2\times10^{12}$m$^{-3}$ during the measurement duration, low elevation cut-off leads to undesirable modulation of the measured spectrum within the bandwidth of interest ($40$-$85$~MHz). We thus recommend monitoring of ionospheric TEC throughout observations with an intention of flagging data during periods with high TEC values.\\

\item To evaluate chromatic effects from more realistic sky and beam models, we have set-up simulations that accept a variety of sky, beam, and ionospheric parameters and produce dynamic spectra of the measured antenna temperature. We have evaluated chromatic effects for the \citet{decosta} sky models, for ideal frequency independent, and simulated LOFAR LBA beams. Using the results of these simulations we have placed limits on the efficiency of spectral basis methods--- algorithms that separate foregrounds from the $21$~cm signal based on the spectral smoothness of foreground and chromatic effects as compared with the $21$~cm signal. In doing so we have shown that even under ideal ionospheric conditions a minimum of $4$ parameters are required to sufficiently describe chromatic effects of the LOFAR LBA beam and the ionosphere, and that a minimum of $\sim 50$\% of the variance in the $21$~cm signal is typically lost due to confusion with foreground and chromatic effects, rendering such a method ineffective. \\

\item We have also evaluated the efficiency of two practical algorithms: (i) \texttt{logpolyfit} that uses polynomial in logarithmic space as basis functions, and (ii) \texttt{svdfit}--- a novel algorithm proposed herein that uses the dynamic part of the antenna temperature spectrum to compute a near-optimal set of basis vectors that may then be used to separate foreground from the time averaged (static part) antenna temperature spectrum. We show that \texttt{logpolyfit} fails as a spectral basis method in our frequency range of interest, but \texttt{svdfit} has potential to succeed in case of a near frequency independent beam. We however conclude that \texttt{svdfit} fails as a spectral basis method in case of the highly chromatic LOFAR LBA. Moreover, dynamic ionospheric effects will only decrease the efficiency of any spectral basis method.\\

\item We conclude that spectral basis methods, though attractive due to their simplicity, do not use many of the strong priors (that almost always exist) from independent measurements of the sky brightness temperature, ionospheric conditions, and the dipole beam, all three of which have a large impact on the dynamic spectrum. As part of future work, we plan to improve upon spectral basis methods for single dipole experiments by modeling the observed data using a full measurement equation. We have recently concluded a pilot project with data from the DE602 LOFAR station (near Garching, Germany), and plan to present results from a measurement equation based modeling of this pilot data in a forthcoming paper. Additionally, embedding the dipole into an array gives dipole-dipole visibilities that provide strong constraints on not only receiver bandpass calibration, but also on all the above factors. Array based measurements/constraints may be an effective way forward for global $21$~cm experiments. Consequently, as part of future work, we also plan to develop a framework that exploits priors derived from dipole-dipole visibilities, to address the challenges posed by large foregrounds and chromatic mixing.  \\

\item While we only address single dipole measurements in this paper, our analysis of chromatic lensing in the ionosphere (see sections \ref{subsec:flayer} and \ref{subsec:leco}) may have implication for interferometric $21$~cm experiments. This is especially important for observations at epochs when the Galactic plane is at low elevations where refraction is at its strongest. Refraction may cause bright radio sources to cross the spectrally varying radio-horizon within the observation bandwidth giving insidious frequency structure in the visibilities. As part of future work, we plan to evaluate these effects on techniques presented by \citet{parsons12} and \citet{vedantham12} that propose to filter out discrete foreground sources in the frequency domain. 
\end{enumerate}

\section*{Acknowledgments}
HKV and LVEK acknowledge the financial support from the European Research Council under ERC-Starting Grant FIRSTLIGHT - 258942. The authors also thank Michel Arts from the Netherlands Institute for Radio Astronomy (ASTRON) for providing the simulated LOFAR LBA beam patterns.

\end{document}